\documentclass[preprint,aps,prd]{revtex4-1}
\usepackage[applemac]{inputenc}
\usepackage{amsmath,amssymb}
\usepackage{graphicx}
\usepackage{float}
\usepackage{hyperref}

\begin{document}
\title{Distorted black holes: nonextensivity in the gravitational wave emission}

\author{H. P. de Oliveira}
\email{henrique.oliveira@uerj.br}
\affiliation{Departamento de F\'{\i}sica Te\'orica - Instituto de F\'{\i}sica A. D. Tavares, Universidade do Estado do Rio de Janeiro, R. S\~ao Francisco Xavier, 524. Rio de Janeiro, RJ, 20550-013, Brazil}



\begin{abstract}
We investigate the interaction between a non-rotating black hole and incoming gravitational waves using the characteristic formulation of the Einstein field equations, framed as a Bondi problem. By adopting retarded time as the null coordinate and recognizing that the final state is invariably a black hole, we show that an apparent horizon forms once sufficient mass accretes onto the black hole. We derive the evolution of the Bondi mass and compute its final value, enabling us to quantify the fraction of the incident mass absorbed by the black hole. Additionally, we establish a nonextensive relation for the absorbed mass as a function of initial parameters, such as the amplitude of the gravitational wave data. 

\vspace{8cm}


This essay received an Honorable Mention in the 2025 Essay Competition of the Gravity Research Foundation.

\end{abstract}

\date{\today}

\maketitle


The golden years of black hole physics convey a period of time from the end of the 50s until the 70s in which notable progress in understanding the main features and properties of black holes. To mention some, we have the stability of the Schwarzschild solution under small perturbations \cite{reggewheeler,vishve,zerilli}, the discovery of the quasinormal modes \cite{vishve2,press} that later served as the final stages of the merger of two black holes, the discovery of rotating black holes \cite{kerr}, the establishment of the uniqueness theorem for black holes \cite{israel}, the thermodynamics \cite{bekenstein} and evaporation of black holes \cite{hawking}. The main ingredients to achieve such a theoretical advance were a complete understanding of the structure of black hole solutions and the investigation of black hole responses under small perturbations by distinct matter fields (scalar, electromagnetic fields, for instance) and gravitational waves. 

The exploration of the nonlinear excitations of a black hole has compelling motivations. A distorted black hole emits gravitational waves that carry out information about its source. Therefore, it represents a formidable theoretical laboratory to explore the dynamics of gravitational waves with the mass extraction from the distorted black hole spacetime. Another nice motivation was the representation of the later stages of two black holes during a merger. For this nonlinear investigation, numerical integration of the resulting field equations was mandatory, implying that numerical techniques needed to be developed.

The first step in investigating distorted black holes was the construction of initial configurations representing a black hole interacting with gravitational waves \cite{abrahams_et_al_92,aninos_et_al_94,brandt_seidel_95_II,camarda_seidel_98,camarda_seidel_99,baker_et_al_2000,brown_lowe_2004,hpoliveira_rodrigues}. Then, several authors explored the resulting dynamical aspects, such as the apparent horizon formation and gravitational radiation emission with the corresponding waveforms \cite{aninos_et_al_94,brandt_seidel_95_II,camarda_seidel_98,camarda_seidel_99,baker_et_al_2000,brandt_seidel_96,brandt_seidel_95_I}. The field equations were presented in the Cauchy formalism in these works but in Refs. \cite{papadopoulos,nerozzi,barreto} the authors studied the same problem with the characteristic initial value problem \cite{bondi,winicour_lrr} focusing on the relevant aspects of generated waveforms.

In the present essay, we have considered an axisymmetric distorted black hole under the lens of the characteristic scheme with outgoing light cone foliations instead of ingoing light cones of Refs. \cite{papadopoulos,nerozzi,barreto}. The novelty here is to focus on determining the final black hole mass after the absorption of part of the gravitational waves that initially surround the black hole. Naturally, the amount of absorbed mass energy depends on the initial strength of gravitational waves. Therefore, we are interested in the relationship between the initial strength of gravitational waves and the amount of absorbed mass.

We start with the metric established by Bondi et al. \cite{bondi} describing an axisymmetric and asymptotic flat spacetimes:

\begin{eqnarray}
	ds^2=-\left(\frac{V}{r}\mathrm{e}^{2\beta}-U^2 r^2 \mathrm{e}^{2\gamma}\right) du^2 - 2\mathrm{e}^{2\beta} du dr 
	- 2 U r^2 \mathrm{e}^{2\gamma} du d\theta + r^2(\mathrm{e}^{2\gamma} d \theta^2 + \mathrm{e}^{-2\gamma}\sin^2 \theta d\varphi^2). \nonumber \\ \label{eq3}
\end{eqnarray}

\noindent Here, $u$ is the retarded time coordinate such that $u=constant$ denotes the outgoing null cones; the radial coordinate $r$ is chosen demanding that the surfaces $(u,r)$ have area equal to $4 \pi r^2$ and the angular coordinates $(\theta,\varphi)$ are constant along the outgoing null geodesics. The metric functions $\gamma, \beta, U$ and $V$ depend on the coordinates $u,r,\theta$ and satisfy the vacuum field equations $R_{\mu\nu} = 0$. 

We express the metric function $V(u,r,\theta)$ as

\begin{equation}
	V(u,r,\theta) = r - 2M_0+S(u,r,\theta),
\end{equation}

\noindent where if $S(u,r,\theta)=0$ together with $\beta=U=\gamma=0$, we recover the Schwarzschild solution with black hole mass $M_0$. For convenience, we defined a new radial coordinate $\eta$ by

\begin{equation}
	r=2M_0(1 + \eta),
\end{equation}

\noindent where the spatial domain $r \geq 2M_0$ now corresponds to $\eta \geq 0$, effectively excising the black hole.  

By noticing that $M_0$ is a natural scale of spacetime, we redefine some of the metric functions:

\begin{equation}
	V \rightarrow \frac{V}{2 M_0},\quad S \rightarrow \frac{S}{2 M_0},\quad U \rightarrow 2 M_0 U, 
\end{equation} 

\noindent and defined two auxiliary functions, $\Gamma$ and $Q$ by

\begin{eqnarray}
	\Gamma &\equiv& (1+\eta) \gamma, \\
	\nonumber \\
	Q &\equiv& (1+\eta)^4 \mathrm{e}^{2(\gamma - \beta)}U_{,\eta},
\end{eqnarray}

\noindent where $Q \rightarrow 2 M_0 Q$ and comma denotes the partial derivative with respect to $\eta$. As a final step to absorb the factor $M_0$, we have to rescale the retarded time $u \rightarrow u/2 M_0$.

The regularity conditions at the symmetry axis require that $\gamma /\sin^2 \theta$ and $U/\sin \theta$ be continuous at $\theta = 0, \pi$. Consequently, the same conditions apply to $\Gamma$ and $Q$. In the numerical code, we adopt normalized variables: $\beta \rightarrow \beta/\sin^4 \theta$, $\Gamma \rightarrow \Gamma/\sin^2 \theta$, $U \rightarrow U/\sin \theta \rightarrow U$, and $Q \rightarrow Q/\sin \theta$.

Before presenting the results, we need to establish the boundary conditions. All the metric functions vanish near the black hole horizon at $\eta=0$. The asymptotic conditions must guarantee the flatness character of the spacetime so that we have adopted the Bondi frame \cite{bondi} that dictates

\begin{eqnarray}
	\Gamma &=& \Gamma_0(u,\theta)+\mathcal{O}(\eta^{-1}),\quad \beta=\mathcal{O}(\eta^{-2}),\quad U= \mathcal{O}(\eta^{-2}), 
	\nonumber \\
	V &=& \eta-2\mathcal{M}(u,\theta) + \mathcal{O}(\eta^{-1}),
\end{eqnarray} 

\noindent where $\mathcal{M}$ is the rescaled Bondi mass aspect. This quantity is crucial for the evaluation of the Bondi mass, $M_B(u)$, whose expression, after taking into account the rescaled metric functions, is given by

\begin{equation}
	M_B(u) = M_0\left(1+\int_0^\pi \mathcal{M}(u,\theta) \sin \theta \, d \theta\right).
\end{equation}

\noindent In the absence of interaction with gravitational waves, $\mathcal{M}=0$, and the Bondi mass reduces to the black hole mass $M_0$. 

The field equations of the Bondi problem are divided into two sets. The first set consists of the hypersurface equations for the metric functions, or hypersurface variables, $\beta, U, Q$, and $V$. These equations take the form $F_{,r} = H_F$, where $F$ represents any of the aforementioned metric functions and $H_F$ generally contains nonlinear terms involving all metric functions \cite{winicour_lrr}. The second set includes the evolution equation for the auxiliary function $\Gamma$, given by $\Gamma_{,ur} = H_{\Gamma}$, where $H_{\Gamma}$ also depends on the metric functions.

We integrate the field equations using a spectral code based on the Galerkin-Collocation method \cite{alcoforado_et_al_2022,new25} that approximates each metric function as a series expansion in specific basis functions — one for the radial dependence and another for the angular dependence. In our case, we use redefined Chebyshev functions for the radial basis under a mapping $\eta=L_0 (1+y)/(1-y)$, $L_0$ being the map parameter \cite{boyd}, to transform the physical domain $0 \leq \eta < \infty$ into the computational domain 
$-1 \leq y \leq 1$. For the angular dependence, we select Legendre functions as the basis functions. The coefficients of the spectral approximations, which depend on the retarded time $u$, are the quantities to be determined. A key advantage of this approach is that it transforms each hypersurface equation into a system of linear algebraic equations for the respective coefficients of $\beta, Q, U$, and $V$. Meanwhile, the evolution equation is approximated as a nonlinear dynamical system governing the coefficients of $\Gamma$.

We start the spacetime evolution with the following initial configuration:

\begin{equation}
	\Gamma(u=u_0,\eta,x) = \epsilon \eta (1-x^2) \mathrm{e}^{-(\eta-\eta_0)^2/\sigma^2},
\end{equation}

\noindent where $\epsilon$ controls the gravitational wave initial strength, $\eta_0$ is the center of the wave distribution and $\sigma$ its width. We set $\eta_0=3$ and $\sigma=1$ in all simulations. 

We translated the above initial data in terms of the spectral coefficients of   $\Gamma$ to feed the corresponding dynamical system, and together with the hypersurface equations, we evolve the spacetime. For this task, we have adopted the Cash-Karp adaptative stepsize integrator \cite{CK} to integrate the dynamical system.

The decay of the Bondi mass is a clear manifestation of the gravitational wave mass extraction. However, since part of the wave atmosphere initially around the black hole falls into it, the final configuration will always be a new black hole with mass $M_0 + \Delta M_{\mathrm{abs}}$, where $\Delta M_{\mathrm{abs}}$ is the mass absorbed by the black hole during the process. Consequently, an apparent horizon always forms, which is signalized when the expansion of the radial null rays,  $\Theta = \mathrm{e}^{-2 \beta}/2M_0(1+\eta)$ vanishes, meaning $\beta \to \infty$ at some point resulting in the break down of the numerical integration. 

\begin{figure}[htb]
	\includegraphics[width=7.5cm,height=6.0cm]{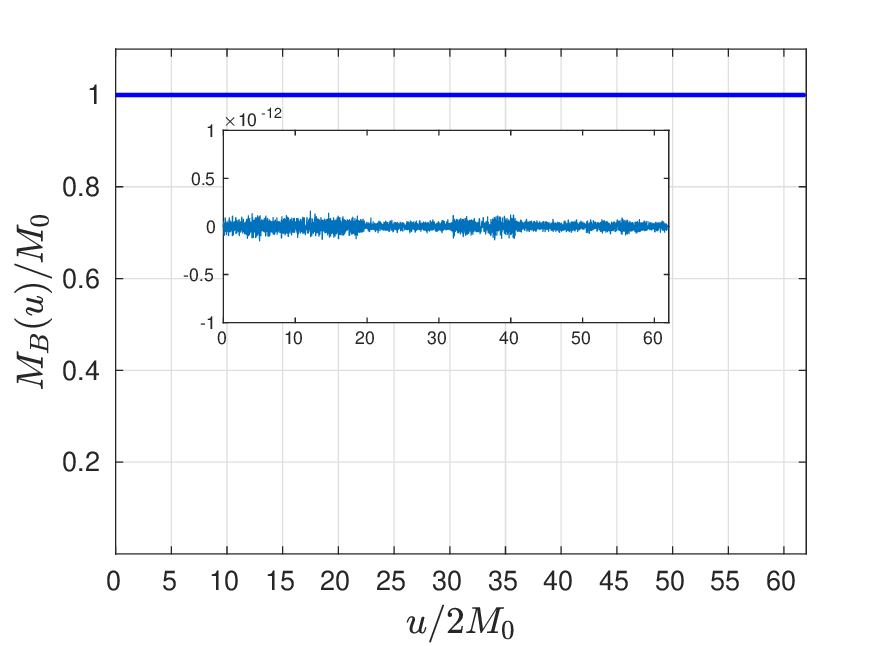}
	\includegraphics[width=7.5cm,height=6.0cm]{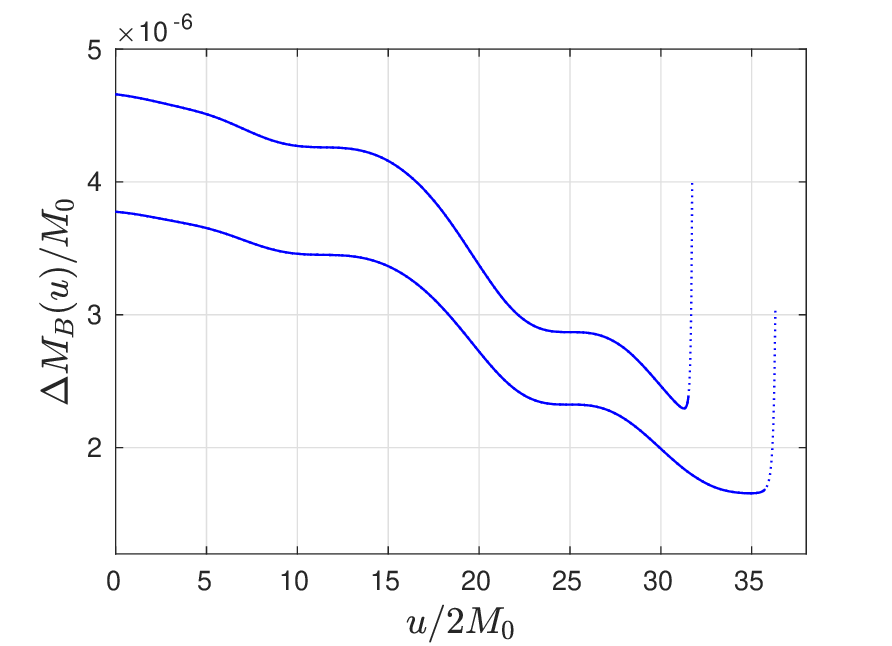}
	\caption{Left panel: evolution of the Bondi mass for $\epsilon=5 \times 10^{-10}$ showing the it does not alter in time. Notice that the linear regime is recovered with $\Delta M_B \approx 0$ numerically. Right panel: decay of $\Delta M_B(u)/M_0$ for $\epsilon=0.0011$ and $\epsilon=0.0009$.}
\end{figure}

From Eq. (8), we define the amount of mass due to the gravitational wave atmosphere around the black hole evaluated at $u$ by

\begin{equation}
	\Delta M_B(u) \equiv M_B(u)-M_0.
\end{equation} 

\noindent \noindent We first illustrated in Figure 1 the linear regime characterized when the perturbations induced by the gravitational waves are negligible, resulting in $\Delta M_B(u) \approx 0$ and leaving the original black hole mass unchanged. In contrast, even for small values of $\epsilon$ (right panel), $\Delta M_B(u)$ decays until the integration stops signalizing the formation of an apparent horizon. 

We are interested in studying the behavior of the total absorbed mass, denoted by $\Delta M_B(u_f)$, as a function of the initial amplitude, $\epsilon$, where $u_f$ is the final retarded time immediately preceding the failure of integration. It is convenient to define the absorbed mass relative to the unperturbed black hole mass, $M_0$, as

\begin{equation}
	\delta M_{\mathrm{abs}} \equiv \frac{\Delta M_B(u_f)}{M_0} =  \frac{M_B(u_f)-M_0}{M_0}.
\end{equation}

\noindent Before presenting our numerical results, we first consider the case of a very small initial amplitude ($\epsilon \ll 1$). From the initial data, we have $\Gamma \sim \mathcal{O}(\epsilon)$, and the field equations dictate the following scaling relations

\begin{eqnarray}
	\beta \sim \mathcal{O}(\epsilon^4)\;\;U \sim \mathcal{O}(\epsilon^2),\;\;S \sim \mathcal{O}(\epsilon^2).
\end{eqnarray}

\noindent Since the mass aspect $\mathcal{M}(u,\theta)$ is derived from the function $S$, we infer that $\Delta M_B(u) \sim \mathcal{O}(\epsilon^2)$. Consequently, the relative absorbed mass satisfies

\begin{eqnarray}
	\delta M_B(u_f) \sim \mathcal{O}(\epsilon^2).
\end{eqnarray}

\noindent This result suggests a power-law relationship between $\delta M_B(u_f)$ and the initial amplitude $\epsilon$, regardless of the specific choice of initial data provided $\epsilon \ll1 $.

We obtained the numerical data by computing the absorbed mass for various values of $\epsilon$. In addition to the initial data (9), we have considered initial data for a pure quadrupole mode 

\begin{equation}
	\Gamma(u=0,\eta,x) = \epsilon\, \mathrm{e}^{-(\eta - \eta_0)^2/\sigma^2}, 
\end{equation} 

\noindent where we have set $\eta_0=1$ and $\sigma=1$, as well as

\begin{equation}
	\Gamma(u=0,\eta,x) = \frac{\epsilon(1-x^2)}{(1+2\eta^2)}.
\end{equation} 

We performed the numerical integration of the dynamical system with appropriate resolution \cite{new25}. Notably, as mentioned before, even for small values of $\epsilon$, $\epsilon \sim \mathcal{O}(10^{-6})$, an apparent horizon forms, leading to a black hole with a small absorbed mass. Increasing the initial amplitudes revealed the following nonextensive law \cite{tsallis,picolli} that best fits the numerical data for any initial data family:

\begin{equation}
	\delta M_{\mathrm{abs}} = K_0\left[1-(1-\alpha \epsilon^\nu)^\mu\right],
\end{equation}

\noindent where $\alpha \approx 0.1408,\,\mu \approx 1.0134$, and $\nu \approx 2.0517$ for \textit{all} initial data families; the proportionality constant $K_0$ depends on the particular initial data set. Figure 2 displays plots of the numerical data (symbols) alongside the above distribution (lines), illustrating the agreement between the theoretical and numerical distributions.

\begin{figure}[htb]
	\includegraphics[width=7.5cm,height=6.0cm]{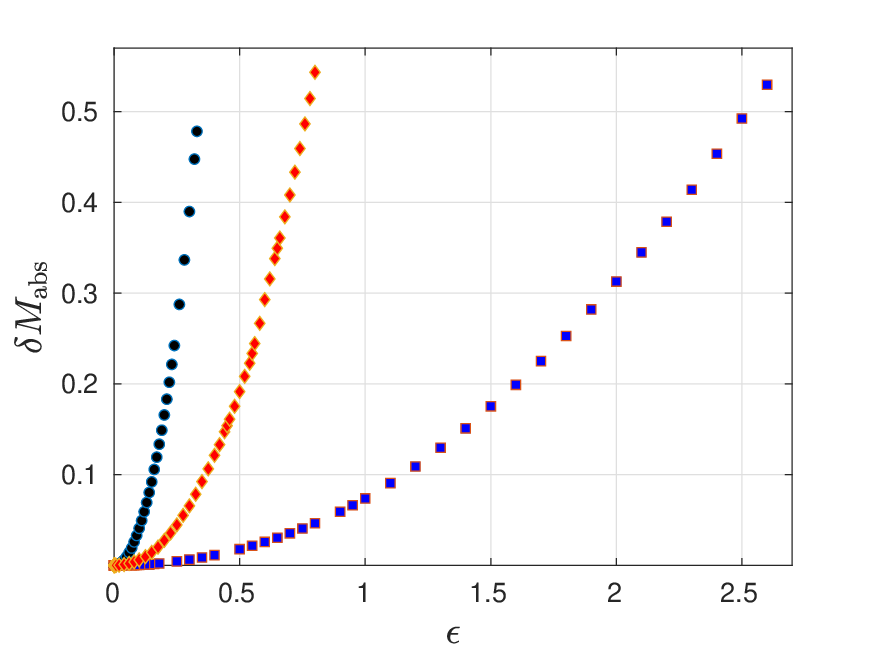}
	\includegraphics[width=7.5cm,height=6.0cm]{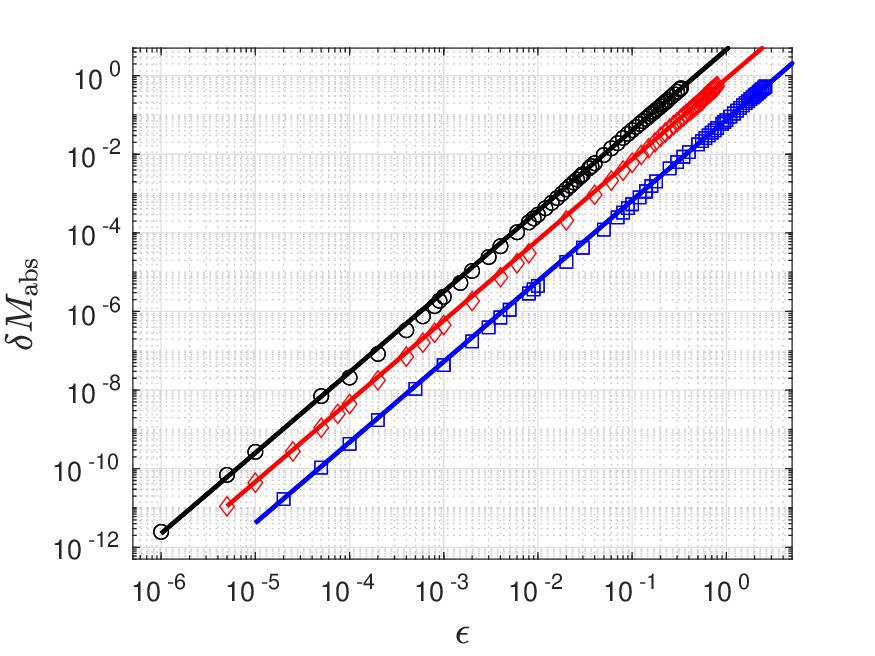}
	\caption{Left panel: absorbed mass $\delta M_{\mathrm{abs}}$ in function of the initial amplitude $\epsilon$ for the initial data families given by Eqs. (9) (circles), (14) (diamonds), and (15) (squares). Right panel: log-log plots of the numerical data (symbols) and the theoretical nonextensive relation (16).}
\end{figure}

Notice that we have a near nonextensive distribution since $\mu$ is close to the unit, and also, for small values of $\epsilon$, we recover the theoretical prediction since $\delta M_{\mathrm{abs}} \approx K_0 \alpha \epsilon^\nu$, with the exponent close to $2$.   

The present results have significant implications. The existence of a unique nonextensive distribution, with scaling law $\delta M_{\mathrm{abs}} \propto \epsilon^\nu$ as a specific case, points to a universality class similar to other scaling laws, such as those associated with the formation of small black hole masses during critical collapse or Kolmogorov scaling in turbulence. Moreover, the universality of the nonextensive distribution suggests a robust and generic mechanism that governs how black holes accrete matter, specifically gravitational radiation. 


We can also interpret our findings as providing insights into the nonlinear stability of black holes in the context of gravitational wave perturbations, revealing a clear selection rule that governs the balance between what is absorbed and what is emitted by the black hole. From the plots in Figure 2, we estimate the maximal increase in mass around the black hole about $0.5 M_0$, meaning a wave atmosphere of gravitational waves with an equivalent mass of about $50\%$ of the original black hole mass $M_0$ for all initial data families. In this regime of strong nonlinear perturbation,  we find that at most $3.6\%$ of the total initial Bondi mass is converted into gravitational waves. As expected, this efficiency is lower than that observed in the realistic merger of two black holes detected by the LIGO consortium \cite{ligo}, where approximately $4.6\%$ of the total initial mass of the two black holes is radiated as gravitational waves. 

\acknowledgments

We thank Conselho Nacional de Desenvolvimento Cient\'ifico e Tecnol\'ogico (CNPq) and Funda\c c\~ao Carlos Chagas Filho de Amparo \`a Pesquisa do Estado do Rio de Janeiro (FAPERJ)
(Grant No. E-26/200.774/2023 Bolsas de Bancada de Projetos (BBP)).

\end{document}